\documentclass{article}
\usepackage{spconf,amsmath,graphicx}
\usepackage[backend=biber, sorting=none]{biblatex}

\usepackage{multirow}
\usepackage{color}
\usepackage{hyperref}

\usepackage [english]{babel}
\usepackage [autostyle, english = american]{csquotes}
\MakeOuterQuote{"}

\title{CONTACT SURFACE AREA: \\
A NOVEL SIGNAL FOR HEART RATE ESTIMATION IN SMARTPHONE VIDEOS}
\name{Sara Fridovich-Keil and Peter J. Ramadge \thanks{*This work was funded by the Department of Electrical Engineering and School of Engineering and Applied Science at Princeton University.}}
\address{Princeton University\\ Department of Electrical Engineering\\ Princeton, NJ, USA}

\begin{document}
\topmargin=0mm
%
\maketitle

\begin{abstract}
We consider the problem of smartphone video-based heart rate estimation, which typically relies on measuring the green color intensity of the user's skin. We describe a novel signal in fingertip videos used for smartphone-based heart rate estimation: fingertip contact surface area. We propose a model relating contact surface area to pressure, and validate it on a dataset of 786 videos from 62 participants by demonstrating a statistical correlation between contact surface area and green color intensity. We estimate heart rate on our dataset with two algorithms, a baseline using the green signal only and a novel algorithm based on both color and area. We demonstrate lower rates of substantial errors ($>10$ beats per minute) using the novel algorithm (4.1\%), compared both to the baseline algorithm (6.4\%) and to published results using commercial color-based applications ($\geq 6\%$). 
\end{abstract}
\section{Introduction}
\label{sec:intro}

Most methods for home measurement of heart rate rely on reflective photoplethysmography (PPG). PPG involves measuring the light reflected by vascularized skin to infer its blood volume \cite{allen2007photoplethysmography}. The premise of this method in the context of smartphone videos is that hemoglobin in the blood absorbs green light, so the intensity of green light reflected by the skin is related to the volume of blood in the skin, which rises and falls with each heartbeat \cite{cui1990vivo}. PPG comes in two varieties, contact (touching the skin) and non-contact (not touching the skin), both of which are implemented in freely available smartphone applications, although contact PPG tends to provide higher accuracy for heart rate estimation \cite{coppetti2017accuracy}. These applications are not considered medical devices and are not subject to accuracy standards for electrocardiograph (ECG) heart rate monitors set forth by the American National Standards Institute/American Association for the Advancement of Medical Instrumentation, which require measurements to be within 5 beats per minute (bpm) or 10\% of the true value, whichever is greater \cite{ansi2002cardiac}. Nonetheless, studies have reported mean absolute errors well within this tolerance for the applications "Instant Heart Rate," "Heart Fitness" \cite{coppetti2017accuracy}, and "Cardiio" \cite{poh2017validation}, when using contact video PPG with a user's fingertip. 

Despite these low mean absolute errors, published accuracy results for contact PPG include a long tail in the distribution of absolute estimation errors. This long tail is a barrier to user trust in smartphone-based heart rate measurement, as the reported heart rate has been shown to differ from the ground truth by more than 10 bpm on at least 6\% of measurements and differ by more than 20 bpm on at least 4\% of measurements \cite{coppetti2017accuracy}.

In this paper, we extend prior methods of video-based heart rate estimation by utilizing a novel signal present in fingertip videos, fingertip contact surface area, to reduce the tail of the absolute heart rate estimation error distribution. We propose a model of the relationship between the contact surface area signal and the pressure in the fingertip, and validate it empirically by showing that contact area correlates with green color, which is used by PPG to estimate heart rate. We use both green color and surface area to estimate heart rate on a dataset of 786 videos collected with concurrent reference heart rate measurements from 62 participants. We demonstrate that heart rate estimation using both color and area achieves comparable mean absolute error and reduces the occurrence of high absolute errors in the distributional tail compared to a baseline implementation using only green color. 

\section{Subjects, Model, and Methods}

\subsection{Study Subjects}
Our study population consisted of 62 adult volunteers; all recruitment materials and study procedures were approved by the Institutional Review Board of Princeton University. Subjects included 37 female and 25 male participants with mean $\pm$ standard deviation ages of 30.2 $\pm$ 15.4 and 27.2 $\pm$ 12.9 years old, respectively. Participants ranged in age from 18 to 64 years old. Our dataset includes a total of 786 fingertip videos (approximately 13 per participant), each with a concurrent reference heart rate measurement. Reference heart rates had a mean $\pm$ standard deviation of 77.1 $\pm$ 13.9 bpm. 

\subsection{Study Procedure}
Reference measurements were taken using an Omron arm cuff, model BP742, which is certified to measure heart rates between 40 and 180 bpm to within $\pm$ 5\% of the true value \cite{omron}. All participants were seated for several minutes prior to data collection, and roughly two thirds were asked to exercise after the third measurement in order to elevate their heart rate transiently. \footnote{Participants with elevated resting blood pressure were not asked to exercise.} Videos were recorded at heart elevation using an iPhone 5 (1080 by 1920 pixels at 30 frames per second) with the flash on to provide continuous illumination. Videos were recorded using the index finger of the arm not used for concurrent reference measurements.

\begin{figure}[t!hb]
\begin{minipage}[b]{1.0\linewidth}
  \centering
  \centerline{\includegraphics[width=0.6\linewidth]{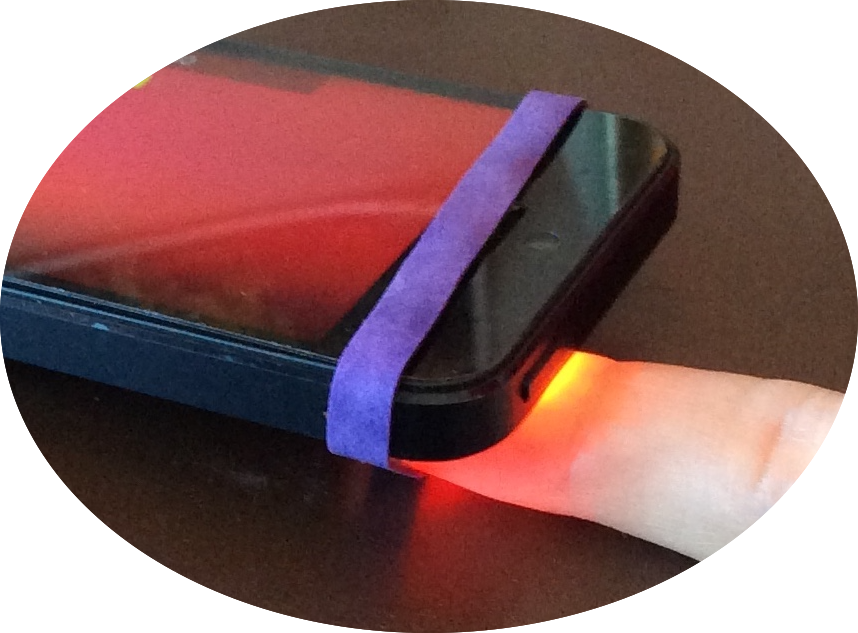}}
\end{minipage}
\caption{Video recording setup.}
\label{fig:fingersetup}
\end{figure}

Fig.~\ref{fig:fingersetup} illustrates the video recording setup. The fingertip was resting on a tabletop nail side down, supporting the weight of the smartphone, thereby maintaining roughly constant force on the fingertip. The fingertip was positioned to cover the flash and a portion of the camera lens, so that the edge of the fingertip contact surface appeared in the video, as shown in Fig.~\ref{fig:ellipseestimation}. A loose rubber band was used to help position the fingertip without restricting blood flow.

\subsection{Model}
\label{sec:model}
We model the fingertip as an elastic ellipsoid, where the contact surface of the fingertip against the smartphone is subject to the relation:
\begin{equation}
area = \frac{force}{pressure} \label{eq1}
\end{equation}
If the finger is held still, the force of the smartphone against the fingertip can be approximated as constant over time. Accordingly, we expect the contact surface area to vary inversely with the pressure inside the fingertip. The volume of blood in the fingertip rises and falls with each heartbeat, implying that the pressure in the fingertip rises and falls accordingly. It follows that both contact surface area and green color intensity vary with blood volume in the fingertip. In Section~\ref{sec:results}, we validate this model by showing a statistical correlation between contact surface area and green color intensity.

\subsection{Heart Rate Estimation}
\begin{figure}[t!hb]
\begin{minipage}[b]{1.0\linewidth}
  \centering
  \includegraphics[width=0.45\linewidth]{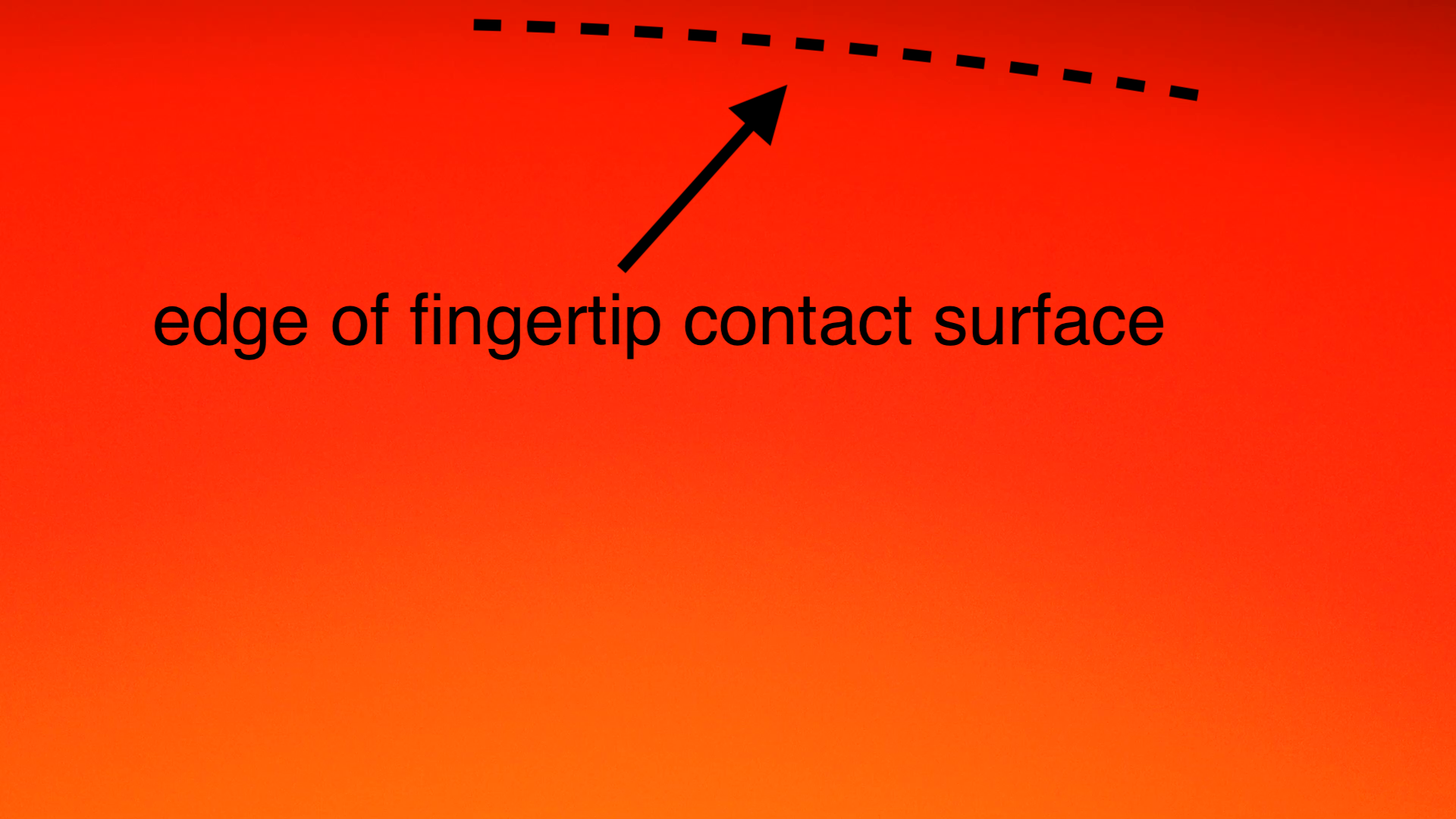}
  \includegraphics[width=0.45\linewidth]{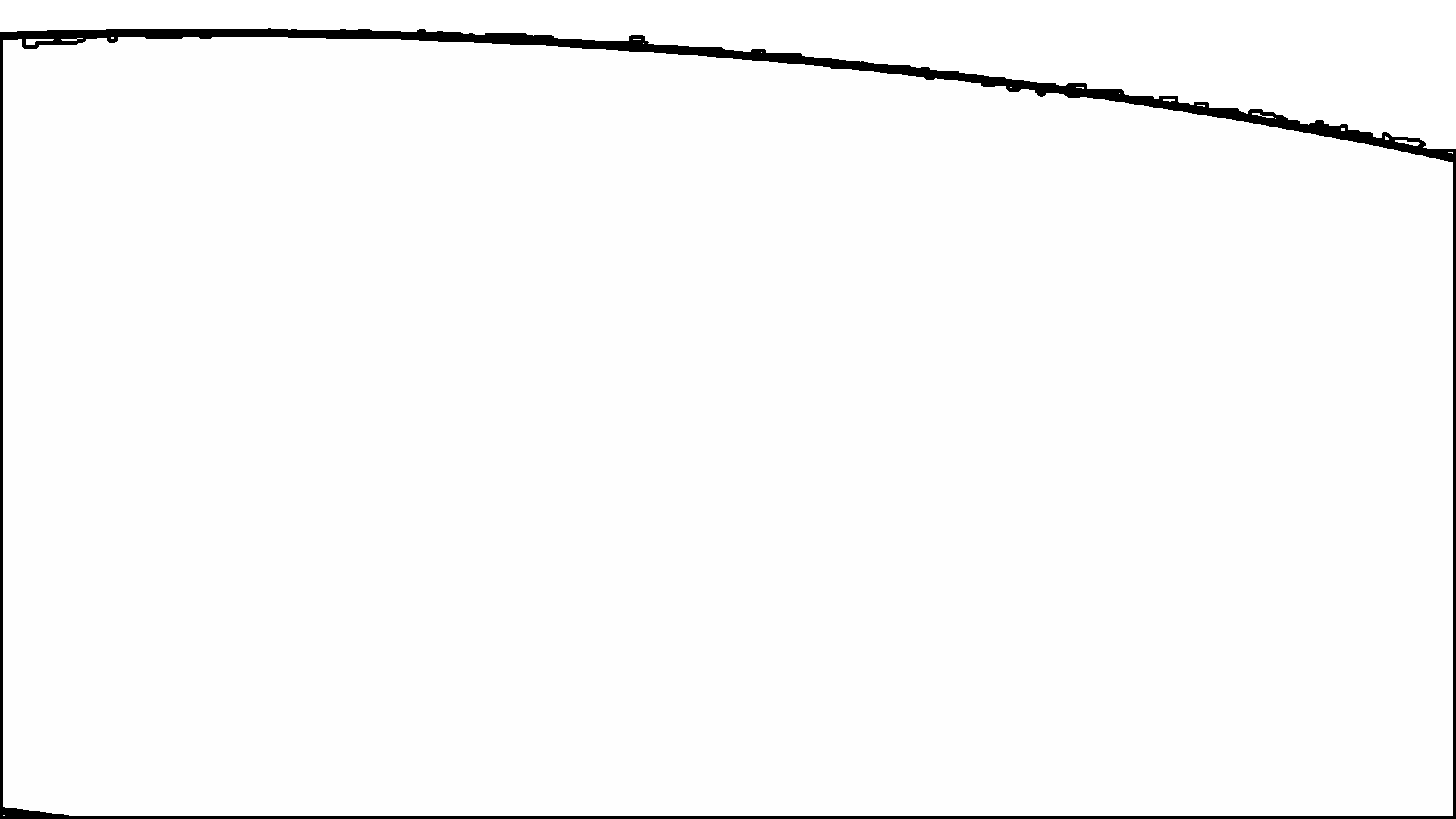}
\end{minipage}
\caption{Raw video frame (left) and estimated ellipse (right).}
\label{fig:ellipseestimation}
\end{figure}

The contact area of the fingertip in each video frame was approximated as an ellipse, as shown in Fig.~\ref{fig:ellipseestimation}. We found the contour of the contact region by applying Otsu's thresholding method \cite{otsu1979threshold} to the red channel, which is robust to glare from the camera flash, and then fitted an ellipse to the contour using Algorithm 3.1 described by Fitzgibbon and Fisher \cite{fitzgibbon1996buyer}. The green color signal was taken as the mean green pixel intensity over all pixels inside the contact area. 

We estimated heart rate for each video using two algorithms: a baseline PPG-based algorithm ("PPG-only") using only green color and a novel algorithm ("PPG-area") combining contact area and green color. Determining the fundamental frequency of these time-domain signals may be viewed as a power spectral density estimation problem, for which classic algorithms exist, e.g. MUSIC \cite{hayes1996statistical}. In this work, we use an ensemble approach combining estimates based on the FFT and group LASSO \cite{yuan2006model} over the FFT coefficients to encourage a sparse frequency representation. The functions in each group were a sine and a cosine at the same frequency, regularized jointly so that the magnitude but \textit{not} the phase was penalized. For each method, the frequency between 30 and 180 bpm with the highest magnitude was taken as the heart rate estimate, and videos where the two estimates differed by more than 5 bpm were deemed excessively noisy and rejected. We randomly separated our 62 participants into training and testing sets, computed the mean squared heart rate estimation error using FFT and group LASSO on the training dataset, and then combined our estimates ($HR_F$ and $HR_L$, respectively) on the testing dataset to minimize mean squared error, under the assumptions that the two heart rate estimates are independent and normally distributed:
\begin{equation}
HR_{Combined} = \frac{\sigma_{L}^2}{\sigma_{F}^2 + \sigma_{L}^2} HR_{F} + \frac{\sigma_{F}^2}{\sigma_{F}^2 + \sigma_{L}^2}HR_{L}  \label{eq2}
\end{equation}
where $\sigma_F^2$ is the mean squared training error using FFT and $\sigma_L^2$ is the mean squared training error using group LASSO. 

PPG-area transformed both contact area and green color into the frequency domain separately using the FFT and group LASSO approaches. Each approach produced a heart rate estimate by taking the frequency between 30 and 180 bpm corresponding to the maximum element-wise product of the area and color frequency coefficient magnitudes. A video was rejected if the maximal product magnitude did not surpass a threshold, or if the estimates produced by FFT and group LASSO differed by more than 5 bpm. The FFT and group LASSO estimates were combined in the same manner as in the PPG-only algorithm.

Further details regarding both PPG-only and PPG-area can be found in the implementation code at \nolinkurl{https://github.com/sarafridov/FingertipVideo}.

\section{Results}
\label{sec:results}

\subsection{Model Validation}

\begin{figure}[b!ht]
\begin{minipage}[b]{1.0\linewidth}
  \centering
  \centerline{\includegraphics[width=0.6\linewidth]{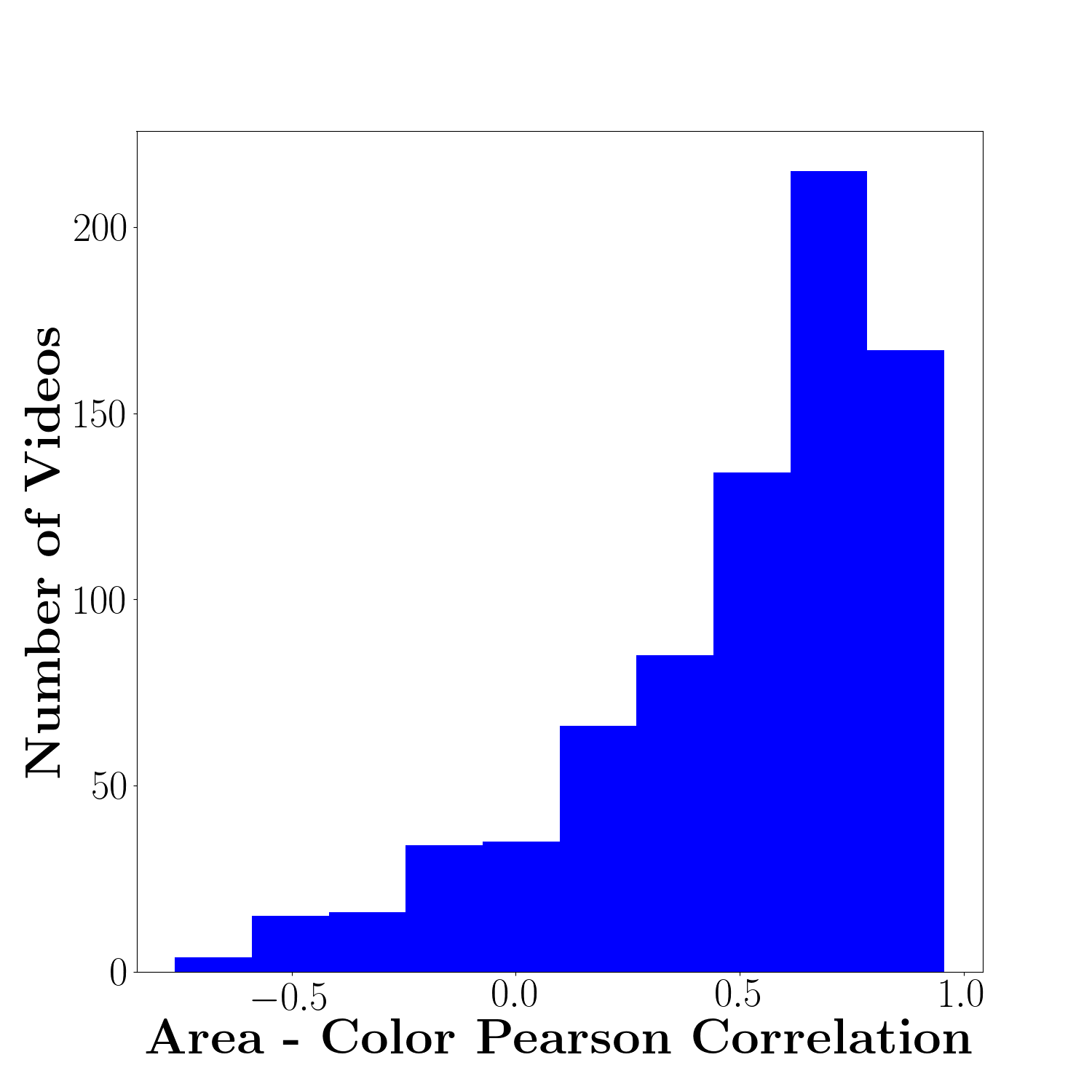}}
\end{minipage}
\caption{Histogram of Pearson correlation coefficients between detrended fingertip ellipse area and green color intensity.}
\label{fig:pearson}
\end{figure}

We validated the model proposed in Section~\ref{sec:model} by computing the Pearson correlation coefficient between contact area and green color in each of the 771 (98\%) videos for which the edge of the fingertip was visible for at least 20 seconds, as shown in Fig.~\ref{fig:pearson}. After removing baseline trends separately in each signal to compensate for finger motion using a method based on that described by Tarvainen \textit{et al.} \cite{tarvainen2002advanced}, the mean $\pm$ standard deviation correlation was 0.50 $\pm$ 0.35, where 62\% of videos showed correlation $>0.5$, median correlation was 0.60, and 11.4\% of videos had negative correlation. 

\begin{figure}[htb]
\begin{minipage}[b]{1.0\linewidth}
  \centering
  \centerline{\includegraphics[width=\linewidth]{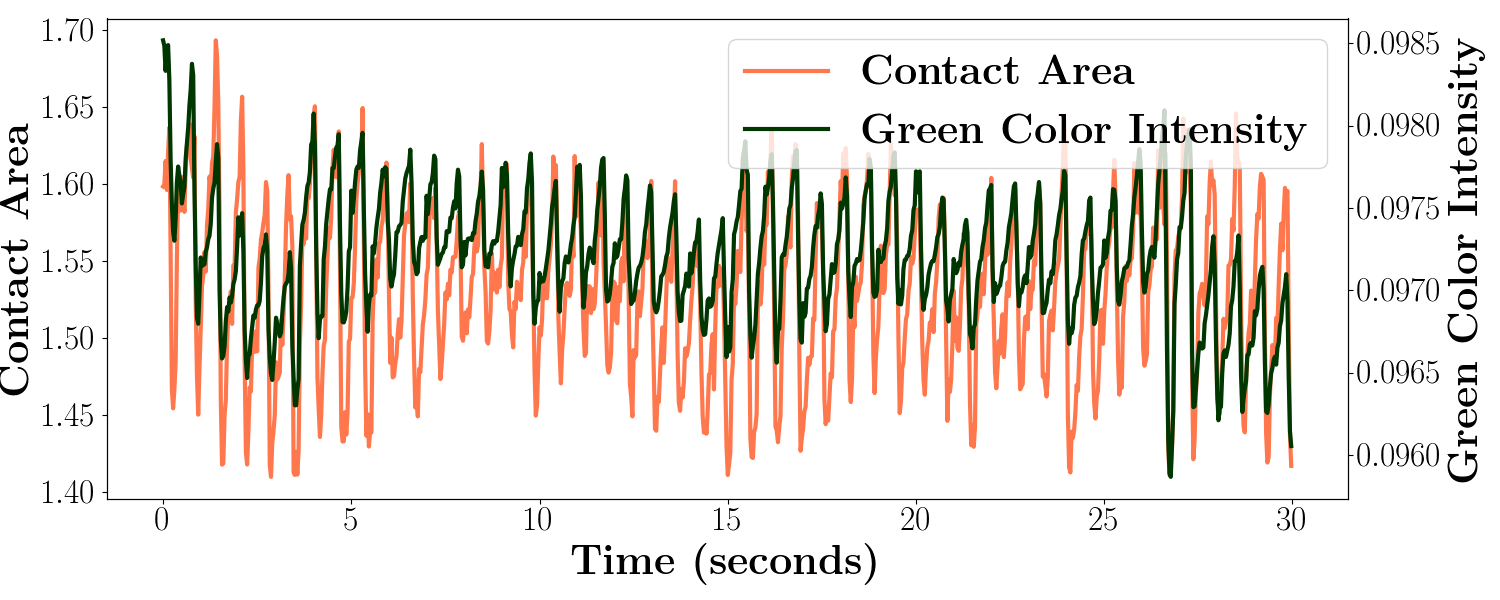}}
  \centerline{\includegraphics[width=\linewidth]{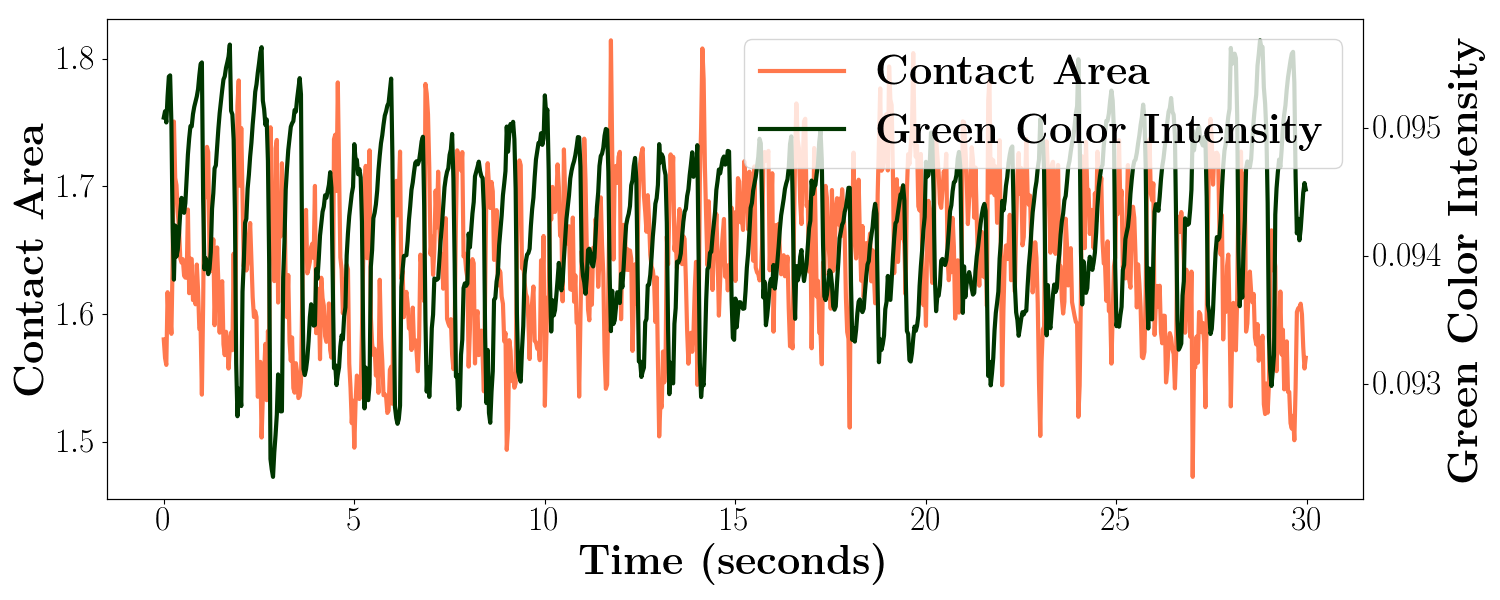}}
\end{minipage}
\caption{Example raw fingertip contact area (normalized to screen size) and green color intensity over time for two videos from different participants.}
\label{fig:rawdataintime}
\end{figure}

Fig.~\ref{fig:rawdataintime} shows the raw area and color time series for two videos. Area and color are in phase and positively correlated in the first video, as in the majority of videos. Area and color are out of phase and negatively correlated in the second video, yet both signals contain clear variation at the same frequency: the heart rate. 

Although both contact area and green color vary with blood volume, the former is transduced optically and the latter mechanically. The precise relationship between blood volume and these quantities likely depends on individual parameters such as skin thickness and elasticity. A more thorough modeling of this relationship may explain why a small proportion of the videos exhibit negative correlation between green color and surface area, as discussed further in Section~\ref{sec:future}. 

\subsection{Improved Heart Rate Estimation with PPG and Area}

\begin{figure}[htb]
\begin{minipage}[b]{1.0\linewidth}
  \centering
  \includegraphics[width=0.49\linewidth]{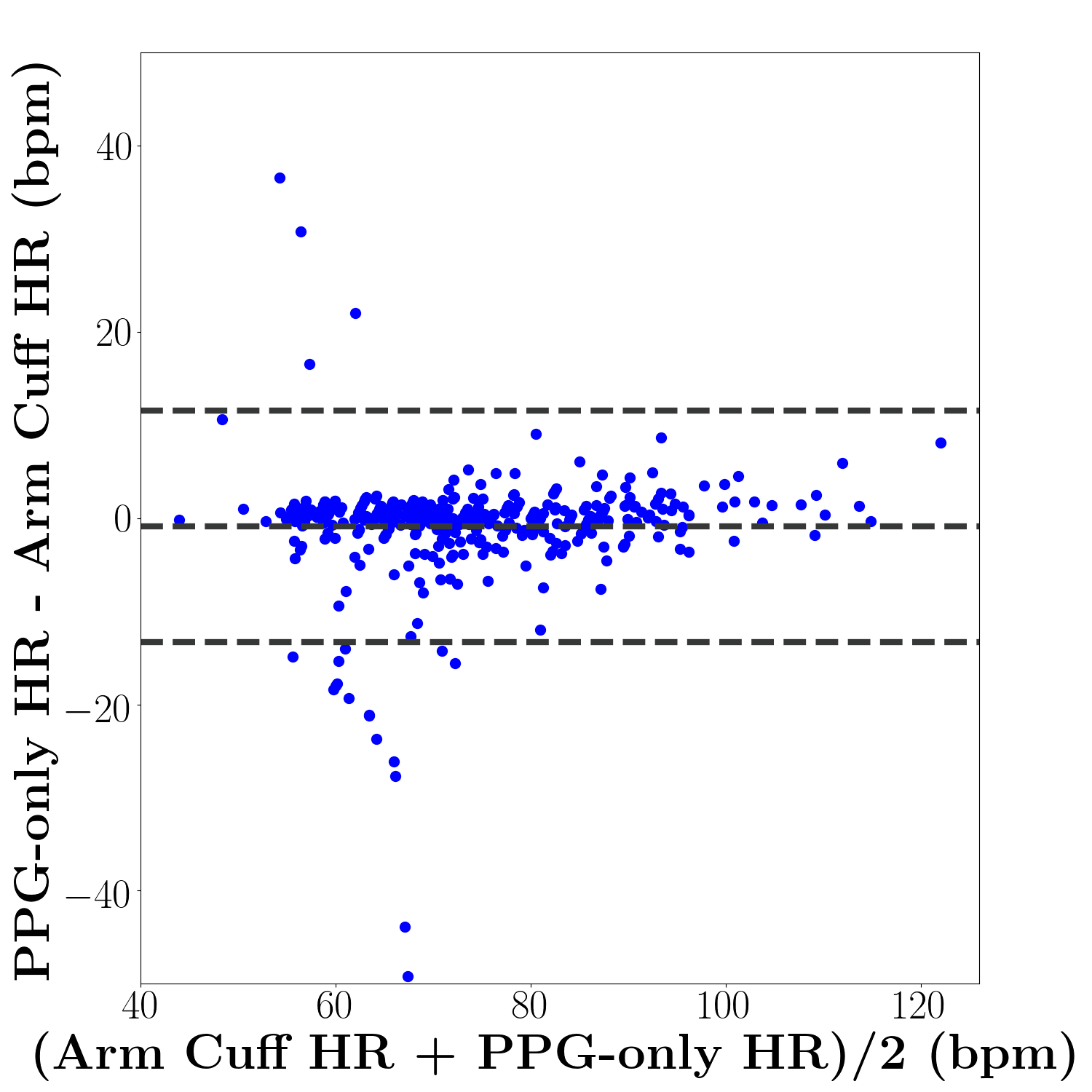}
  \includegraphics[width=0.49\linewidth]{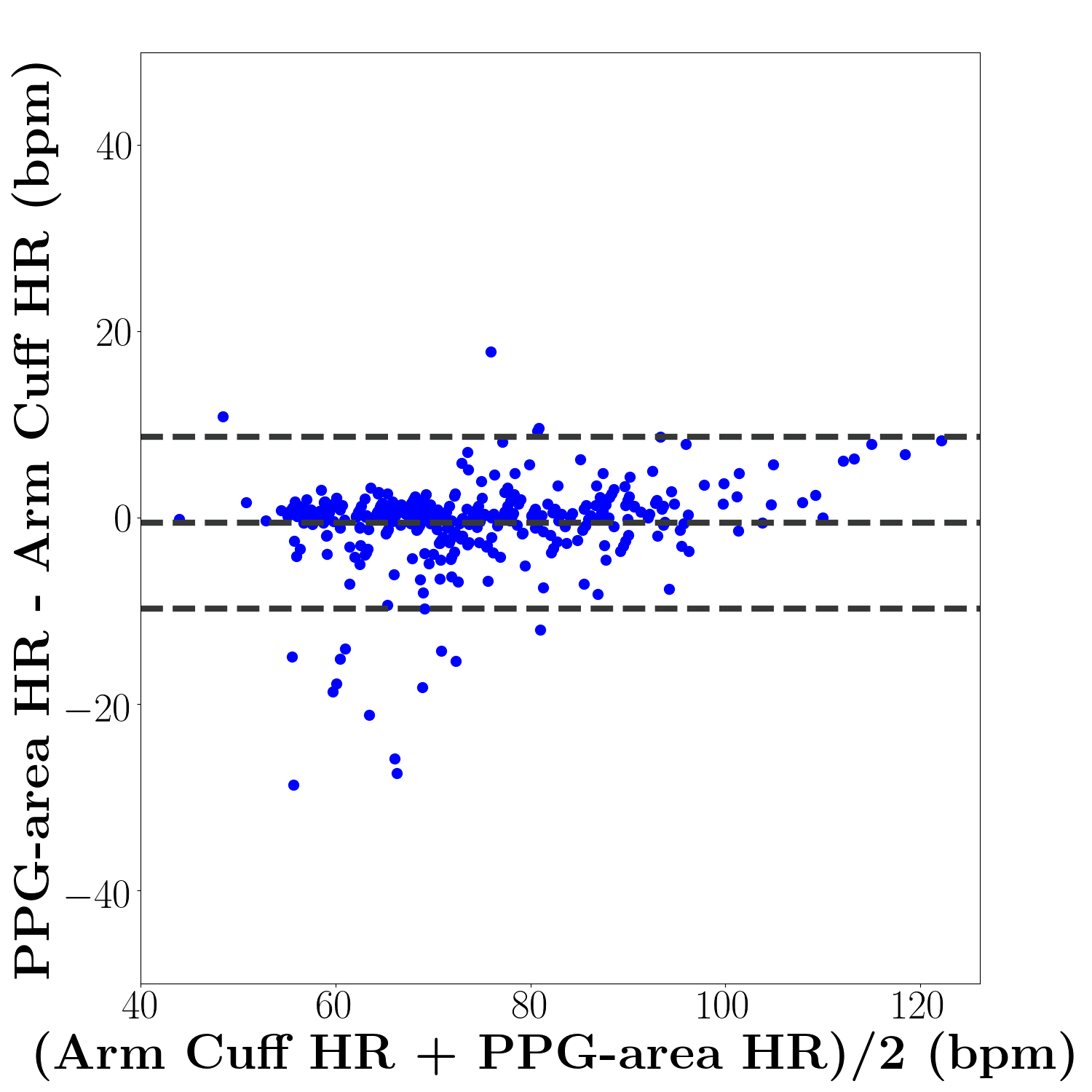}
\end{minipage}
\caption{Bland-Altman plots showing the accuracy of PPG-only (left) and PPG-area (right) heart rate estimation compared to reference heart rate measured using an arm cuff. The dotted lines represent the mean and 95\% limits of agreement for the difference between estimated and reference heart rates, showing tighter agreement for PPG-area.}
\label{fig:bland-altman}
\end{figure}

Table~\ref{table:hrresults} presents key attributes of the two algorithms. PPG-area rejected a slightly higher percentage of the testing videos, but achieved higher accuracy in all metrics compared to PPG-only. PPG-area had lower mean absolute testing error and lower rates of substantial errors, as well as stronger agreement with the reference measurements. Bland-Altman plots \cite{bland1986statistical} comparing these algorithms are shown in Fig.~\ref{fig:bland-altman}.

\begin{table}[htb]
\begin{center}
\begin{tabular}{|c||c|c|}
\hline
 & PPG-only & PPG-area\\
\hline \hline
Videos accepted (\%) & \textbf{93.5} & 91.25 \\
\hline
Mean abs. testing error & \multirow{2}{*}{2.91 $\pm$ 0.29} & \multirow{2}{*}{\textbf{2.60 $\pm$ 0.21}}\\
$\pm$ standard error (bpm) &  & \\
\hline
Videos with abs. testing & \multirow{2}{*}{6.42} & \multirow{2}{*}{\textbf{4.11}}\\
error $>10$ bpm (\%) & & \\
\hline
Videos with abs. testing & \multirow{2}{*}{2.67} & \multirow{2}{*}{\textbf{1.10}}\\
error $>20$ bpm (\%) & & \\
\hline
Videos with abs. testing & \multirow{2}{*}{0.53} & \multirow{2}{*}{\textbf{0.00}}\\
error $>40$ bpm (\%) & & \\
\hline
Pearson's $r$  & 0.89  & \textbf{0.94}  \\
($p < 0.00001$) & ($n=374$) & ($n = 365$)\\
\hline
Mean bias (bpm) & -0.84 & \textbf{-0.51} \\
\hline 
95\% limits of agreement  & \multirow{2}{*}{(-13.27, 11.59)} & \multirow{2}{*}{\textbf{(-9.71, 8.69)}} \\
with arm cuff (bpm) & & \\
\hline
\end{tabular}
\caption{Heart rate estimation results using PPG-only and PPG-area, where error is with respect to reference heart rate measurements using the arm cuff on the testing dataset.}
\label{table:hrresults}
\end{center}
\end{table}

The mean absolute testing errors for both PPG-only and PPG-area were within the reported accuracy of our reference arm cuff instrument. PPG-only had greater than 10 bpm absolute error on nearly the same percentage of measurements (6\%) as did "Heart Fitness," the most accurate PPG-based application for which comparable data were available \cite{coppetti2017accuracy}. PPG-only had absolute error more than 20 and more than 40 bpm at slightly lower rates (2.6\% vs. 4\% and 0.5\% vs. 2\%, respectively) than the most accurate commercial applications for which these statistics were available \cite{coppetti2017accuracy}. PPG-area achieved a lower rate of absolute error more than 10 bpm (4\% versus 6\%), 20 bpm (1\% versus 4\%), and 40 bpm (0\% versus 2\%) \cite{coppetti2017accuracy}. PPG-area likewise achieved lower rates of substantial error compared to PPG-only, as well as stronger agreement ($r = 0.94$ versus $r = 0.89$) with the reference heart rate. 

\section{Discussion}

In this study, we proposed a model describing the relationship between fingertip contact surface area and pressure, which is a physical indicator of blood flow. We validated this model by demonstrating a positive correlation between contact surface area and green color intensity, the signal commonly used in PPG-based heart rate estimation. We also demonstrated that the contact area contains additional information that is not present in the green color, and that including the area signal in addition to green color can apply this information to reduce the incidence of high absolute errors in heart rate estimation. 

\subsection{Limitations}

This study was subject to limitations in both dataset and data analysis. Our study population included predominantly young adults (18-22 years old). Although many participants were asked to exercise briefly, nearly all of our reference heart rates were below 120 bpm. A second limitation is that although participants were asked to hold their measurement finger still and flat against the tabletop, many accidentally moved their finger slightly during video recording. Further, our reference heart rate measurements were limited by the accuracy of the arm cuff device. 

We designed features and algorithms intended to approximate those used by commercial PPG-based applications, with the addition of the area signal, but we could not directly replicate those algorithms because they were proprietary. As such, we were able to draw direct comparisons between our PPG-only and PPG-area algorithms, but there may be specific features of our algorithms that differed from the algorithms running in the various PPG-based commercial applications. Nevertheless, our PPG-only algorithm achieves error rates on our dataset comparable to the most accurate commercial PPG-based applications \cite{coppetti2017accuracy}, suggesting that PPG-only captures the salient features of the commercial algorithms.

\subsection{Future Work}
\label{sec:future}

The fingertip contact surface area model introduced in this paper may be developed further in both applications and theory. 

Possible applications of this area signal include video-based measurement of blood pressure and hematocrit. If our area model holds, systolic blood pressure should be inversely proportional to the minimum contact surface area over a heartbeat, while diastolic blood pressure should be inversely proportional to the maximum contact surface area over a heartbeat. Fingertip contact surface area in combination with green color intensity may also enable video-based hematocrit measurement, which is the volume fraction of red blood cells in the blood and is used to diagnose and monitor anemia. The area signal is influenced primarily by blood volume, while the green color signal is influenced primarily by the amount of hemoglobin present in the red blood cells. 

Future theoretical work will be devoted to refining the model proposed here to account for the complex physical and optical properties of the fingertip. One promising next step would be to utilize the on-device accelerometer to estimate net forces on the fingertip over time and account for them algorithmically. A more precise modeling of the effects of fingertip tissue elasticity may additionally explain the distribution we observed in the correlation between the color and area signals recorded from different participants. The model could be further improved by considering the impact of the inverse relationship in the time domain between pressure and area on the area signal in the frequency domain. 

These future directions present promising opportunities for public health and medical research.

\section{REFERENCES}
\label{sec:refs}

\printbibliography[heading=none]

\end{document}